\documentclass[11pt]{article}
\usepackage{moriondarch,epsfig}

\def\be{\begin{equation}}
\def\ee{\end{equation}}
\def\bea{\begin{eqnarray}}
\def\eea{\end{eqnarray}}
\def\ba{\begin{array}} 
\def\ea{\end{array}}
\def\bc{\begin{center}}
\def\ec{\end{center}}

\def\ghost#1{}

\def\simge{\mathrel{%
   \rlap{\raise 0.511ex \hbox{$>$}}{\lower 0.511ex \hbox{$\sim$}}}}
\def\simle{\mathrel{
   \rlap{\raise 0.511ex \hbox{$<$}}{\lower 0.511ex \hbox{$\sim$}}}}

\def\Journal#1#2#3#4{{#1} {\bf #2}, #3 (#4)}

\def\NPB{{\em Nucl. Phys.} B}
\def\PLB{{\em Phys. Lett.}  B}

\def\PRD{{\em Phys. Rev.} D}

\def\mco{\multicolumn}

\def\ra{\rightarrow}

\def\ko{K^0}

\def\be{\begin{equation}}
\def\ee{\end{equation}}
\def\bea{\begin{eqnarray}}
\def\eea{\end{eqnarray}}

\begin{document}
\vspace*{1.5cm} 
\title{{\large \ \ LIGHT \,DARK \,MATTER} \,\footnote{\ Proc.~39$^{\rm th}$ Rencontres
de Moriond \,``Exploring the Universe'',
La Thuile, Italy (march-april 2004) http://moriond.in2p3.fr/J04/proceedingsm04.html, july 2004.
\hfill LPTENS-04/35} \medskip}

\author{P. FAYET \bigskip}

\address{Laboratoire de Physique Th\'eorique de l'ENS (CNRS),\\
24 rue Lhomond, 75231 Paris Cedex 05, France \bigskip\bigskip}

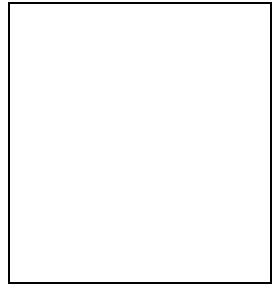
\maketitle\abstracts{
We show how {\it \,light\,} spin-$\frac{1}{2}$ or spin-0 particles may be acceptable
Dark Matter candidates, provided they annihilate sufficiently strongly through
new interactions, such as those induced by a new {\it light neutral spin-1 boson\,} $U$.
The corresponding interaction is {\it \,stronger than weak interactions\,} at lower energies,
but {\it weaker\,} at higher energies.
\\[1.5mm]
Annihilation cross sections of ({\it axially coupled\,}) spin-$\frac{1}{2}$  Dark Matter particles,
induced by a $U$ vectorially coupled to matter, are the same
as for spin-0 particles. In both cases, the
cross sections $\ (\sigma_{ann}\,v_{rel}/c)\,$ into $\,e^+e^-$
automatically include a $v_{dm}^{\,2}$ suppression factor,
as desirable to avoid an excessive production
of $\gamma$ rays from residual Dark Matter annihilations.
We also relate Dark Matter annihilations
with production cross sections in $\,e^+e^-$ reactions.
As for spin-0, light spin-$\frac{1}{2}$  Dark Matter particles
annihilating into $e^+e^-$
could be responsible for the bright \,511 keV
$\gamma$ ray line observed by INTEGRAL from the galactic bulge.
\\[1.5mm]}

\section{Introduction}

\label{sec:intro}

What is (non-baryonic) Dark Matter made of\,? The familiar, although still unobserved, neutralinos of supersymmetric theories\,?
What could be the other possibilities\,? Could Dark Matter particles be light\,?
If so, how could they annihilate\,? May be through some new interactions,
such as those induced by a new light spin-1 boson $U$,
that would be sufficiently strong for this purpose\,?
But if so, how could they have remained unnoticed\,?
And how could such Light Dark Matter particle annihilations manifest\,?
Could they be responsible for the bright \,511\, keV
\hbox{$\gamma$ ray} line recently observed by the INTEGRAL satellite from the galactic bulge\,?

\vspace{1mm}

We adopt here the usual point of view, according to which Dark Matter is made of new neutral massive particles, not directly observed.
Such particles have to annihilate sufficiently,
otherwise their relic density would be too large.
Therefore weakly-interacting massive neutral particles,
taken as possible Dark Matter candidates, should not be too light.
\vspace{1mm}

Supersymmetric extensions of the Standard Model \cite{ssm}
naturally provide weakly-interacting neutral particles \cite{sigmaphot},
stable as a result of $R$-parity conservation.
Spin-$\frac{1}{2}$ photinos, or more generally neutralinos,
taken as possible Dark Matter candidates, should then be sufficiently heavy
\cite{goldberg}.
In any case, given the still unsuccessful hunt for superpartners, most notably at LEP,
the lightest neutralino (LSP) of Supersymmetric extensions of the Standard Model
is generally believed to be heavier than about $\,\sim 30\,$ GeV.

\vspace{1mm}
Still, their may be room for other, non-conventional possibilities, such as those classified as in \ \cite{class}\, according to the mass and interaction strength of Dark Matter particles,
which determine whether or not they are still relativistic at the time of their decoupling.
More specifically, Light Dark Matter candidates
having sufficiently ``strong'' interactions
could remain coupled until they become non-relativistic and annihilate
(cf. the left part of ``Region II'' in \ \cite{class}),
with annihilation reactions freezing out at $\,T_F=m_{dm}/x_F$,
provided the interactions responsible for these annihilations be strong enough.
But, can this really make sense\,?

\vspace{1mm}

Indeed, how could a light (annihilating) Dark Matter particle possibly exist, without leading to
a too large relic density\,?
At first, in any case, it should have {\it no significant direct coupling to the $Z$ boson},
otherwise it would have been produced in $Z$ decays at LEP.
Despite that, it would have to {\it annihilate sufficiently\,}
-- and in fact, much more strongly than through ordinary weak interactions
-- otherwise its relic energy density would be too high\,!
Can this happen at all, and what could then be the new interactions
responsible for Light Dark Matter annihilations\,?

\vspace{1mm}
We have explored in \cite{bf} under which conditions
a light \hbox{spin-0} particle
could be a viable Dark Matter candidate.
In the most interesting case
the new interactions responsible for its annihilations are mediated by a new neutral spin-1
gauge boson $U$, light and very weakly coupled,
as introduced long ago \cite{pfu}.
The spin-1 induced Dark Matter annihilation cross section
into $e^+e^-$ ($\,\sigma_{ann} \,v_{rel}/c$)
\,also includes, naturally, a $v_{dm}^{\,2}$ low-energy suppression factor, as desired to avoid an excessive production
of $\,\gamma\,$ rays originating from the residual annihilations of Dark Matter
particles (if lighter than $\sim$ 100 MeV) \cite{bes}.

\vspace{1mm}
This applies as well to (Majorana or Dirac) spin-$\frac{1}{2}$ particles \cite{fay}.
Their annihilation cross sections into $f\bar f$ fermion pairs
through the exchanges of a $\,U$ boson have, also in this case, the desired  $v_{dm}^{\,2}$ suppression factor
at threshold, provided the $U$ boson is {\it \,axially\,} coupled
to Dark Matter particles, and
{\it \,vectorially\,} coupled to
{\it \,matter\,} fermions.
This last feature is, in any case, also necessary to avoid a problematic axionlike behavior
of its longitudinal polarization state~\cite{pfu}.
The annihilation, at threshold, of a
$\,C=+\,$ state (with $\,J=S=0$) \,into a $\,f\bar f\,$ state
with $\,C'=(-)^{(L'+S')}= +\,$, \,through a $\,C$-violating interaction,
is forbidden by charge conjugation.
This ensures that $\,\sigma_{ann}\,v_{rel}\,$
has the appropriate $\,\propto\,v_{dm}^{\,2}\,$ behavior,
automatically suppressing (by a factor $\,\approx 10^{-5}\,$)
the late annihilations of non-relativistic relic Dark Matter particles.

\vspace{1mm}
Furthermore, the annihilation cross sections
of \hbox{spin-$\frac{1}{2}$} and spin-0 Dark Matter particles are, in this case,
given by {\it \,exactly the same expressions}. (Majorana or Dirac)
\hbox{spin-$\frac{1}{2}$} particles then turn out to be acceptable
Light Dark Matter (LDM) candidates, as well as spin-0 particles.

\vspace{1mm}
One crucial feature is that
the new interactions mediated by the $U$ boson should actually be
``not-so-weak'' (at lower energies and relatively to weak interactions) \,--
i.e. $<\!\sigma_{ann} \,v_{rel}/c\!> \ \,\approx $ 4\,--\,5 picobarns for a Majorana particle,
or 8\,--\,10 pb   \,for a Dirac one or a complex scalar,
so as to ensure for sufficient annihilations
of light Dark Matter particles, whatever their spin~\cite{fay}.
More precisely, the new $\,U$-mediated Dark-Matter/Matter interactions
will be {\it stronger than ordinary weak interactions} but only {\it at lower energies},
when weak interactions are really very weak.
But {\it weaker at higher energies}, \,at which they are damped by
$\,U$ propagator effects (for $s$ or $|q^2| > m_U^{\,2}$), when weak-interaction cross sections,
still growing with energy like $s$, become important.
The smallness of the $\,U$ couplings to ordinary matter,
as compared to $\,e$, \,by several orders of magnitude,
and the resulting smallness of $\,U$ amplitudes compared to electromagnetic ones,
\,then accounts for the fact that these particles have not been observed yet.

\vspace{1mm}
We have indicated in may 2003
that ``a gamma ray signature from the galactic centre at low energy could be
 due to the existence of a light new gauge boson'', inducing
annihilations of Light Dark Matter particles into $\,e^+e^-$ \ \cite{bf}.
The observation, a few months later, by the satellite INTEGRAL
of a bright \,511\, keV \,$\gamma$ ray line from the galactic bulge \cite{integral},
requiring the presence in this region of a rather large number of annihilating positrons,
may then be viewed as originating from
such Light Dark Matter particle annihilations \cite{betal}.
Indeed spin-0, but just as well spin-$\frac{1}{2}$ particles,
could then be responsible for this bright \,511\, keV line,
which does not seem to have an easy interpretation in terms
of known astrophysical processes involving conventional physics~\cite{fay,bfsi}.

\section {Dark Matter decoupling and relic density}
\label{sec:relic}

Let us evaluate the (relatively large) cross sections
needed to obtain a suitable relic abundance of Light Dark Matter particles, of mass $\,m_{dm}$ \cite{fay}.
Before that, we note that for spin-$\frac{1}{2}$ particles $\chi$
the interactions responsible for the annihilations
may be written, in the local limit approximation,
as effective four-fermion interactions
$\,{\cal L}\,\approx \,G\ \,\bar \chi ... \chi $  $ \bar f ... f\,$.
The corresponding cross sections, proportional to $\,G^2$, \,scale like $\,m_{dm}^{\,2}$.
Such {\it \,fermionic\,} particles that would annihilate through exchanges of (charged) heavy
{\it bosons\,} of masses $\,\simge m_W$ cannot be light, since their annihilation cross sections
would be too small.

\vspace{1mm}
To estimate what cross sections are needed for a correct relic abundance,
we express that the annihilation rate
$\,\Gamma =n_{dm}\!<\!\sigma_{ann}\,v_{rel}\!>\,$
and expansion rate $\,H\,$
are approximately equal when Dark Matter annihilation reactions freeze out
at a temperature $\,T_F=m_{dm}/x_F$
(with $\,x_F\,\simeq \,16\,$ to $\,23 \,$ for a 1 MeV to 1 GeV particle).
We concentrate here on light masses
$\,m_{dm}\simle 1$ GeV.

\vspace{3mm}
{\large $\bullet$}\ \ For {\,10 MeV  $\simle m_{dm}\simle 1$ GeV}, \,the freeze-out occurs at $T_F$
(between $\,\simeq \,.6\,$ and \,50  MeV) after most muons have annihilated, but not electrons yet.
The surviving particles get diluted by
the expansion of the Universe, proportionally to $\,T^3$,
\,with an extra factor $\,4/11\,$ corresponding to the subsequent annihilation
of $\,e^+e^-$ pairs into photons, so that the relic density of Dark Matter particles
may now be expressed as
\be
\label{densite0}
n_{\circ \,dm}\ \,=\ \frac{4}{11}\ \,\frac{T_{\circ\gamma}^{\ 3}}{T_F^{\,3}}\ \ n_{dm}\ \ ,
\ee
$T_{\circ\gamma}\simeq \,2.725 \ \hbox{K}\,$
being the present photon temperature.
We denote by $\,N_{\circ\,dm} = (2) \ n_{\circ\,dm}\,$ the total number density of
Dark Matter (particles~+~antiparticles), with the factor 2 present only for
non-self-conjugate particles.
The freeze-out equation $\ \Gamma\,\simeq\,H\,$,
\be
\label{gamma=h0}
n_{\circ\,{dm}}\ \frac{11}{4}\ \frac{T_F^{\,3}}{T_{\circ\gamma}^{\,3}}\
<\sigma_{ann}\,v_{rel}>\ \,\simeq\
1.66\ \sqrt{\hbox{\small $g_*\!=\frac{43}{4}$}}\ \ \frac{T_F^{\,2}}{m_{Pl}}\ ,
\ee
fixes the relic energy density
\be
\label{rhodm0}
\ba{ccl}
\rho_{dm}  \,\simeq\,
\ (2)\ \,n_{\circ\,{dm}}\ m_{dm}
&\simeq& \displaystyle  (2)\ \,\frac{4}{11}\ \,1.66\ \sqrt{\hbox{\small $g_*\!=\frac{43}{4}$}}
\ \, x_F\ \frac{T_{\circ\gamma}^{\,3}}{m_{Pl}}\ \
\frac{1}{<\sigma_{ann}\,v_{rel}>}
\vspace{2mm}\\
&\simeq& \displaystyle  (2)\ \,\frac{x_F}{20}\ \ \
\frac{4.2\ 10^{-56}\ {\rm GeV}^2}{<\sigma_{ann}\,v_{rel}>}\ \ .
\ea
\ee
Dividing by the critical density $\,\rho_c/h^{2}\simeq 1.054\ 10^{-5}$\ GeV/cm$^3$,
\,we get the density ratio $\,\Omega_{dm}\,h^2$
\be
\label{omegah20}
\frac{\Omega_{dm}\ h^2}{.1} \ \simeq \
(2)\ \ \frac{x_F}{20}\ \
\frac{2\ \,10^{-36}\ {\rm cm}^2}{<\sigma_{ann} \,v_{rel}/c>}\ \ \ .
\ee

\vspace{1mm}

There is also an expected increase of the required averaged cross section by a factor $\,\approx 2$,
\,for  \linebreak
$\,<\!\sigma_{ann}v_{rel}\!>\,$ \,behaving at threshold like
$\,v_{dm}^{\,2}.\,$
(Indeed the
later annihilations that would still occur below the temperature $\,T_F$ given by eq.\,(\ref{gamma=h0})
are further inhibited by this $\,v_{dm}^{\,2}\,$ factor, preventing the Dark Matter density from reaching the equilibrium value corresponding to this $\,T_F$.)
Altogether, obtaining the right amount of Dark Matter ($\,\Omega_{dm}\,h^2\simeq \,0.1\,$)
\,requires typically
\be
\label{sigv}
<\sigma_{ann}v_{rel}/c >
\ \ \simeq\ (2)\ \,4 \ \hbox{pb}\ \ ,
\ee
when $\,<\sigma_{ann}\,v_{rel}>\,$  behaves like $\,v_{dm}^{\,2}$,
the factor \,2\, being associated with non self-conjugate Dark Matter particles.

\vspace{2mm}
{\large $\bullet$}
\ For lighter particles, the evaluation of the relic abundance gets modified, without leading, however, to drastic effects in the cross sections \cite{fay}.
For particles lighter than about 2 to 3 MeV,
that would decouple (at $\,T_F\simle .15$ MeV)
\,after most electrons have annihilated,
the dilution factor $\,4/11\,$ is no longer present in eqs.\,(\ref{densite0}-\ref{rhodm0}).
The $\,g_*\,$ at Dark Matter freeze-out (no longer 43/4) may be expressed in terms of the neutrino temperature as
$
\,g_*\, \simeq\, 2\,+\, \frac{7}{8}\,(2\times 3)$ $\left.\frac{T_\nu^{\ 4}}{T^4}\right. _F$
(which would be $\,\simeq 3.36\,$ in the standard model).
The neutrino contribution to $\,g_*\,$, \,however, is no longer the same.
Dark Matter particles annihilating after neutrinos decouple
(normally at $\,T\approx 3.5\,$ to $\,2\,$ MeV)
would also heat up the photon gas
as compared to neutrinos, so that the resulting $T_\nu$ would be less
than the usual $\,(4/11)^{1/3}\,T_\gamma\,$, \,resulting in {\it a lower contribution of neutrinos\,} to the $\,g_*\,$ at $T_F$\,.
If a significant fraction of Dark Matter annihilations were to occur
after neutrino decoupling but before the $\,n/p$ ratio freezes out,
this could allow for less primordial helium than in the standard model\,!
While light masses $\,m_{dm}\simle\,2\,$ MeV, however, are found to be disfavored as they would severely disturb the BBN concordance, slightly larger masses could in fact improve it \cite{sr}.
In any case, for $\,m_{dm} \simle \,2\,$ MeV,
the required cross section gets slightly increased
(from the absence of the 4/11 factor but with lower values of
$\,g_*\,$ and $\,x_f$)
by $\,\simeq$ 20\,\%, up to about (2) times 5 pb --
with no spectacular difference expected when $m_{dm}$ grows from \,2\, to 10 MeV or more.

\vspace{2mm}
{\large $\bullet$}\ \,Altogether {\it \,for cross sections behaving like} $v_{dm}^{\,2}\,$,
the required
$\,\hbox{$<\sigma_{ann}\,v_{rel}/c>$}\,$'s at freeze-out
are of the order of \,4 to 5\,  picobarns for a self-conjugate (Majorana) Dark Matter particle,
or \,8 to 10\, pb  for a complex scalar, or Dirac particle, as
summarized below \cite{fay}. We do not consider real self-conjugate spin-0 particles, as Bose statistics does not allow for
the desired $P$-wave annihilation.

\vspace{2mm}

\begin{center}

{\footnotesize
Table 1: Estimates of the annihilation cross sections \hbox{$\,<\sigma_{ann}\,v_{rel}/c>\,$}
at freeze out
\,required for a correct relic
\vspace*{-1mm}

abundance \,(\,$\Omega_{dm}\,h^2\simeq\,.1\,$).}

\vspace{3mm}

\begin{tabular}{|c|c|c|}
\hline && \vspace{-2mm} \\
\ \ Spin-$\frac{1}{2}$ Majorana \ \   &\ \  Spin-$\frac{1}{2}$ Dirac\ \  & \ \ Spin-0 \ \
\\ [1mm]
{\small (\,$\chi$\,)} & {\small ($\,\psi$\,)} &
{\small \ \ \, (\,$\varphi\,$ complex\,)} \ \ \, \\
\vspace{-3mm} && \\ \hline
\vspace{-2mm}&& \\
4 -- 5\ pb & 8 -- 10\ pb & \ 8 -- 10\ pb
\vspace{-1mm}\\
&& \\ \hline
\end{tabular}
\end{center}

\vspace{4mm}

\noindent
Since
$\ G_{\!F}^{\,2}\,(1\,\hbox{GeV})^2\,/2\,\pi\,
\simeq \,10^{-38}$ cm$^2$, \,or \,$10^{-2}$  pb,
cross sections (at freeze-out) of weak interaction order are,
for light masses $\,m_{dm} \ll$ GeV,
\,by far too small for a correct relic abundance.
Significantly larger annihilation cross sections are needed,
requiring {\em new types of interactions}~\cite{bf,fay}, if Dark Matter is to be made of such light particles.

\vspace{2mm}
This corresponds roughly, for present annihilations of residual
Dark Matter particles having a velocity $\,v_{dm}\approx 3\ 10^{-3}$ the velocity
at freeze-out in the primordial Universe ($\,\simeq \,.4\ c\,$),
\,and cross sections $\,<\sigma_{ann}\,v_{rel}>\ \propto v_{dm}^{\ 2}\,$,
to
\be
<\sigma_{ann}\,v_{rel}/c>_\circ\ \  \simeq \ \,(\,4\,\ \hbox{to}\ 10\,)\ 10^{-5}\
\hbox{pb}\ \ .
\ee
This is the right order of magnitude for
light Dark Matter particle (in the $\,\simeq\,$ MeV range)
annihilations to be at the origin of the \,511 keV $\gamma$
ray signal observed by INTEGRAL from the galactic bulge\, \cite{integral,betal,fay,bfsi}.

\vspace{2mm}
Given INTEGRAL results, we tend to favor rather light Dark Matter masses,
so as to maximize the total number of $\,e^+$ produced in their annihilations, and therefore the observable signal
expected from a given Dark Matter energy density.
And, also, to avoid these $\,e^+$ and $\,e^-$ from Dark Matter annihilations, initially produced with an energy close to $\,m_{dm}$, \,having too much energy dissipated in $\gamma$ rays,
before $e^+$ can form positronium and annihilate, possibly
leading to the bright 511 keV $\,\gamma$ ray line.

\section{Spin-0 annihilations through heavy fermion ex\-chan\-ges}

Spin-0 Dark Matter particles \,($\varphi$)\,
having Yukawa interactions
coupling ordinary quarks and leptons $f$ to heavy fermions $F$
such as {\it \,mirror fermions\,} (as inspired by $N=2$ extended supersymmetric
and/or higher-dimensional theories\ \cite{mirror})
may have relatively large annihilation
cross sections,
behaving as the inverse of the {\it \,squared masses\,}
of the exchanged mirror fermions.
The low-energy effective Lagrangian density responsible for their
pair-annihilations into $\,f\bar f\,$
may indeed be written as an effective dimension-5 interaction,
\be
\label{leff2}
{\cal L}\ \ \approx \ \
\frac{C_l\,C_r}{m_F}\ \ \varphi^*\varphi\ \ \overline{f_R}\,f_L\,
+\ \hbox{h.c.}\ \ ,
\ee
$C_l$ and $C_r$ denoting the Yukawa couplings to the left-handed and right-handed fermion fields,
respectively. The resulting annihilation cross section at threshold,
of the type
\be
\label{sigmamf}
\sigma_{ann}\,v_{rel}\ \ \approx\ \ \frac{C_l^2\,C_r^2}{\pi\ m_F^2}\ \ ,
\ee
(for {\it \,non-chiral\,} couplings, i.e. $\,C_l\,C_r\,\neq0$),
is largely independent of the Dark Matter mass; it
can be quite significant and take the appropriate values,
{\it even for light \hbox{spin-0} Dark Matter particles} \cite{bf}.

\vspace{2mm}

However, in the absence of a $P$-wave suppression factor $\,\propto\,v_{dm}^{\,2}$
one runs the risk, at least for lighter Dark Matter particles
($\,\simle 100$ MeV$/c^2$), of too much $\gamma$ ray production due to
residual annihilations of Dark Matter particles \cite{bes}
(unless there is an asymmetry between Dark Matter particles and antiparticles).

\section{Exchanges of a new light spin-1 gauge boson $U$}

\label{sec:U}

It is thus preferable to consider annihilations induced through the virtual production
of a new light neutral spin-1 gauge boson $\,U$ \,\cite{bf,pfu,fay}.

\subsection{Effective Lagrangian density for spin-0 annihilations}

Spin-0 Dark Matter annihilations may be described, in the local limit approximation,
by an effective Lagrangian density involving the product of the Dark Matter ($\varphi$)
and quark and lepton ($f$) contributions to the $U$ current, i.e.
\be
\label{effphi}
{\cal L}\ \ =\ \ \frac{C_U}{m_U^{\,2}}\ \ \
\varphi^*\ i \stackrel{\leftrightarrow}{\partial_\mu}\!\varphi\ \
(\,f_V\ \bar f\,\gamma^\mu f\ +\ f_A\ \bar f\,\gamma^\mu\gamma_5 f\,)\ \ ,
\ee
$C_U$ and $f_V$ and/or $\,f_A$ denoting the couplings
of the new gauge boson $U$
to the \hbox{spin-0} Dark Matter field  $\,\varphi\,$
and the matter fermion $f$ considered, respectively.
The $\,f_V$ coupling (i.e. in fact the product $\,C_U f_V$) corresponds to an interaction invariant under Charge Conjugation,
while for the $f_A\,$ coupling (i.e. $\!C_U f_A$) \,this interaction would be $\,C$-violating.

\subsection{Restrictions on axial couplings of the $\,U$ to ordinary matter.}
\label{subsec:limax}

These axial couplings $f_A$ of the new light gauge boson $U$  to matter fermions $f$, however,
are likely to be absent \,-- or at least should be sufficiently small --\, otherwise they would lead to a generally-unwanted axionlike behavior
of the light spin-1 $\,U$ boson. Let us concentrate for a while on what would happen in the presence of such axial couplings ($f_A$) of the $U$ boson.

\vspace{2mm}

The longitudinal polarisation state of a light (ultrarelativistic) $U$ boson,
with polarisation vector $\epsilon^\mu_{\,L}\,\simeq \,k^\mu_{\,U}/m_U$,
\,and coupled proportionately to $\,f_{V,A}\ \epsilon^\mu_{\,L}\,\simeq\,(f_{V,A}/m_U)\ k^\mu_{\,U}\,$,
\,would then behave very much
as a spin-0 axionlike particle, having {\it \,pseudoscalar\,} couplings to quarks and leptons
\be
\label{couplage}
\hbox {\framebox [4.8cm]{\rule[-.45cm]{0cm}{1.1cm} $ \displaystyle {
f_{P\,q,l}\ =\ 2\ f_A\ \ \frac{m_{q,l}}{m_U}\ \ ,
}$}}
\ee
as soon as there is a (non-conserved) axial part in the
quark-and-lepton contribution to the $U$ current.
Indeed $\,k^\mu_{\,U}\,$ acting on an {\it \,axial\,} matter current $\,\bar f\,\gamma^\mu\gamma_5\,f$
resurrects an effective pseudoscalar coupling to
$\,\bar f\,\gamma_5\,f$ with an extra factor $\,2\,m_f\,$ in its coefficient, which leads to
the effective pseudoscalar coupling (\ref{couplage}), as shown in the second paper of Ref. \cite{pfu}
(while $\,k^\mu_{\,U}\,$ acting on a {\it \,vector\,} matter current
$\,\bar f \gamma^\mu f$ gives no such contribution).

\vspace {2mm}

There, the axial couplings $f_A$ were expressed in terms of the extra $U(1)$ gauge coupling $g"$
as $\frac{g"}{4}$,
in the simplest case of a universal axial coupling
of the new gauge boson $U$ to quarks ans leptons (obtained for equal v.e.v.'s for the two Higgs doublets \footnote{Separately responsible for the down-quark and charged-lepton masses
\,(through $\,<\varphi_d>\ =v_1/\sqrt 2$), \,or up-quark masses
\,(through $\,<\varphi_u>\ =v_2/\sqrt 2$),
\,as in supersymmetric extensions of the standard model \cite{ssm},
\,from which these theories originate.},
or $v_1\!=\!v_2\,$, i.e. $x=1$). One also has
(with $r=1$, i.e. no extra Higgs singlet introduced, and the extra $U(1)$ symmetry
spontaneously broken at the electroweak scale),
$\,2\,f_A=\frac{g"}{2}= \,2^{\frac{1}{4}}\,G_F\,\!^\frac{1}{2}\,m_U
\simeq\,4\ 10^{-6}\ m_U(\hbox{MeV}) $. Expression (\ref{couplage}) then
reconstructs exactly the usual pseudoscalar coupling $\ 2^{\frac{1}{4}}\,G_F\,\!^\frac{1}{2}\ m_{q,l}$
\,of a \hbox{spin-0} axionlike particle (the one ``eaten away'' by the light spin-1 $\,U$ boson) to quarks and leptons.  This is phenomenologically unacceptable, as no such axionlike particle has been observed.

\vspace{2mm}

More generally, however, we can introduce as in \,\cite{pfu} an extra Higgs singlet
(with, possibly, a large v.e.v. so that the extra $U(1)$ symmetry would then be broken
``at a large scale'' $F$), and the two Higgs doublet v.e.v.'s are not necessarily equal (their ratio being denoted by $x$).
The effective pseudoscalar coupling (\ref{couplage}) would then read:

\vspace{-4mm}

\be
\label{couplage2}
\hbox {\framebox [15.5cm]{\rule[-.4cm]{0cm}{1cm} $ \displaystyle {
f_{P\,q,l}\,=\,2\ f_{A}\ \ \frac{m_{q,l}}{m_U}\ =\ \ 2^{\frac{1}{4}}\ G_F\,\!^\frac{1}{2}\ m_{q,l}\
\times\
\left\{\
\ba{l}
\, r\,x \ \ \ \ \ \ \ \hbox{\small for \ $u,\, c,\, t$ \ quarks,} \vspace{2mm}\\
r/x \ \ \ \hbox{\small for \ $d,\, s,\, b \ $ quarks \,and \,$e,\, \mu,\, \tau$ \ leptons.}
\ea\right.
}$}}
\ee

\vspace{2mm}

This may now be acceptable if the ratios
$\,f_A/m_U$ (inversely proportional to the extra $U(1)$ symmetry-breaking scale
$\,F$) are sufficiently small
(as it would happen for  $F$ at least slightly above or well above the electroweak scale),
as in the ``invisible  $U$ boson'' or analogous ``invisible axion'' mechanisms \cite{pfu}.
The values of $\,f_{V,A}/m_U$ that we may like to consider (cf. Sec. \ref{sec:value})
to generate sufficient annihilations of Light Dark Matter particles into $\,f\bar f$ pairs
(in practice $e^+e^-$, or $\,\nu\bar\nu$), \,however, could be larger than what would be acceptable for
the ratios $\,f_A/m_U$, unless of course the Dark Matter coupling $\,C_U$ is taken to be sufficiently large. (Note, in addition, that no direct effective $\,U\,\gamma\,\gamma$ coupling is to be expected here, and that the $U$ boson would not decay into $\,\gamma\gamma$, in contrast to an axion.)

\vspace{2mm}
In particular, from the results of $\,\psi\,$ and $\,\Upsilon\,$ decay experiments, sensitive to the decays $\,\psi\to\gamma\,+\,U\,$, $\,\Upsilon\to\gamma\,+\,U\,$ (with the $U$ having invisible decay modes, into Dark Matter particle or $\,\nu\bar\nu$ pairs), we get limits
on the axial couplings of the $U$ to the $c$ and $b$ quarks.
These axial couplings may be expressed, using the language of Refs. \cite{pfu,psi},
as $f_A\,\simeq \,2\ 10^{-6}\ m_U(\hbox{MeV}) \times (rx\ \,\hbox{or} \ \,r/x)\,$ \,for the $\,c\,$
and $\,b\,$ quarks, respectively (cf. eq.\,(\ref{couplage2})).
$\,r\ll 1\,$ corresponds to an extra $U(1)$ symmetry broken
``at a very high scale'' $F$ much larger than the electroweak scale, so that the axionlike effects of the new gauge boson $U$ would then get quasi ``invisible''.
We then get the limits $\,rx<.75\,$ and $\,r/x<.4\,$ from the $\psi\to\gamma\,+\,U\,$
and $\Upsilon\to\gamma\,+\,U\,$ decays, respectively \footnote{We assume here that the $U$ decays preferentially into unobserved neutrals,
as it would happen for a $\,U$ coupling to Dark Matter $\,C_U\,$ significantly larger
than couplings to ordinary matter fermions $f$. Otherwise the limits of (\ref{limabc}) should be slightly weakened.},
which may be turned into the following restrictions on the axial couplings of the $c$ and $b$ quarks,
\be
\label{limabc}
\hbox {\framebox [12cm]{\rule[-.4cm]{0cm}{1cm} $ \displaystyle {
f_{A\,c}\ <\ 1.5\ \,10^{-6}\ m_U\hbox{(MeV)}\ \ ,\ \ \ \
f_{A\,b}\ <\ 0.8\ \,10^{-6}\ m_U\hbox{(MeV)}\ \ .
}$}}
\ee
This may be remembered approximately, at least in the latter case, as
\be
\label{limaq}
\frac{f_{A\,q}^{\ 2}}{m_U^{\,2}}\ <\ \frac{1}{10}\ \,G_F\ \ .
\ee

Further, and potentially stronger, restrictions on the coupling $f_{A\,q}$ may also be obtained
from a careful analysis of searches for $\,K^+\to\pi^+ \,+$ unobserved $U$ boson,
provided one is sufficiently confident about the possibility of estimating reliably
the decay rate in terms of $f_{A\,q}$ and $m_U$~\cite{pfu}.
As an illustration if we consider that the branching ratio $\,B\,(K^+\!\to\pi^+\,U)$
ought to be $\,\simge\,10^{-8}\ r^2/x^2$, and combine this with the experimental limit
 $B\,(K^+\to\pi^+ \,+$  unobserved $U$)\ $\simle .6\ 10^{-10}$ (resp. $10^{-10}$) \,for $m_U\simle$ 70 (resp. 100) MeV~\cite{kpiu}, we get
$\,r/x\simle \,.1$, \,corresponding to
\be
\label{limas}
f_{A\,s}\,\ \simle\ \,2\ \,10^{-7}\ m_U\hbox{(MeV)}\ \ ,
\ee
which may be remembered as $\,f_{A\,s}^{\ 2}/m_U^{\,2}\ \simle\ (1/300)\ G_F\,$!

\vspace{3mm}
A similar analysis may be done for the $\,U$ contribution to the anomalous magnetic moments of
the charged leptons, with appropriate care due to the possibility of cancellations between (positive) vector and (negative) axial contributions, of opposite signs.
In the limit in which $m_U$ is small compared to $\,m_\mu$,
\,one gets from from $f_{V\mu}$ and $f_{A\,\mu}$ the two contributions
\be
\label{amu0}
\delta a_\mu  \ \simeq  \
\displaystyle \frac{f_{V\mu}^{\ 2}}{8\,\pi^2}\ -\
\displaystyle \frac{f_{A\,\mu}^{\ 2}}{4\,\pi^2}\ \ \frac{m_\mu^{\,2}}{m_U^{\,2}}\ \ .
\ee
\vspace{-1mm}
\noindent
If, to keep things simple, we were to disregard the $f_{V\mu}$ contribution
to $\delta a_\mu\,$, we would get \cite{pvU}
\be
\delta_A \,a_\mu\ \simeq\
-\ \frac{f_{A\,\mu}^{\ 2}}{4\,\pi^2}\ \frac{m_\mu^{\,2}}{m_U^{\,2}}\
\,\simeq\,-\,\frac{G_F\,m_\mu^{\,2}}{8\,\pi^2\sqrt2}\ \ \frac{r^2}{x^2}\ \
\simeq\,1.17\ 10^{-9}\ \ \ \frac{r^2}{x^2} \ \
\ee
(as for the exchange of a pseudoscalar axionlike particle),
which implies $\,r/x <1.5\,$, \,corresponding to
\be
f_{A\,\mu}\ <\ 3\ 10^{-6}\ m_U\hbox{(MeV)}\ \ ,
\ee
if we want the extra (negative) $f_{A\,\mu}$ contribution to $a_\mu$ to be, by itself, $\,\simle$
(2 to 3) $10^{-9}$ in magnitude. This bound, approximately expressed as
\be
\frac{f_{A\,\mu}^{\ 2}}{m_U^{\,2}}\  <\ G_F\ \ ,
\ee
\vspace*{-4mm}

\noindent
may of course be relaxed in the presence of a vector coupling
$f_{V\mu}$ sufficiently large compared to $\,f_{A\,\mu}\,$,
as shown by eq.\,(\ref{amu0})~\footnote{Furthermore if $\,m_U$ is large compared to $\,m_e$,
we also have:
\vspace*{-1mm}
$$
\delta a_e \ \simeq \ \displaystyle
\frac{1}{12\,\pi^2}\ \frac{m_e^{\,2}}{m_U^{\,2}} \
\left(\, f_{V\,e}^{\ 2}\,-\,5\,f_{A\,e}^{\,2}\,\right)
\ \ ,
$$
which generally turns out to be less constraining
than the corresponding limits from the muon $\,g-2$; cf. eqs.~(\ref{amu},\ref{amu2}) in Sec. \ref{sec:value}. This may also be reexpressed, using eq.~(\ref{couplage2}), as \cite{pvU}
$$
\ba{c}
\delta a_e \ \simeq \ \displaystyle\
\frac{f_{A\,e}^{\ 2}}{4\,\pi^2}\ \frac{m_e^{\,2}}{m_U^{\,2}}\
\left(-\,\frac{5}{3}\, +\, \frac{1}{3}\ \frac{f_{Ve}^{\,2}}{f_{A\,e}^{\,2}}\,\right)\
\simeq\ \,
\frac{G_F\,m_e^{\,2}}{8\,\pi^2\sqrt2}\ \
\left(-\, \frac{5}{3}\, +\, \frac{1}{3}\ \frac{f_{Ve}^{\,2}}{f_{A\,e}^{\,2}}\,\right)
\ \ \frac{r^2}{x^2}\ \ .
\ea
$$
}.

\vspace{2mm}
Furthermore, atomic physics parity-violation experiments also provide strong constraints on the
coupling constant product $\,f_{A\,e}\,f_{Vq}$, \,such that $\,f_{A\,e}\,f_{Vq}/m_U^{\,2}$
should be less than a small fraction of $G_F$ \cite{pvat}. \,This may also be interpreted as pointing towards a purely vectorial coupling of the new gauge boson $U$ to ordinary matter \,-- unless of course its couplings to matter are taken to be sufficiently small (which again would require the $U$ coupling to Dark Matter $\,C_U$ to be large enough).

\vspace{2mm}
Altogether, we shall generally stick, at least for simplicity, to purely {\it vector}
couplings $\,f_V$ of the $U$ boson to ordinary matter. This is also motivated by the fact that this occurs naturally in a number of $\ SU(3)\times SU(2)\times U(1)\times
\hbox{extra-}U(1)\,$, \,or $\ SU(5) \times \hbox{extra-}U(1)\,$,
\,gauge extensions of the Standard Model (cf. the third paper in Ref. \,\cite{pfu}),
in which the matter couplings of the $U$ boson turn out to be given by a (conserved)
linear combination of the $\,B$, $L$
\,(or $\,B-L$, in grand-unified theories) and electromagnetic currents.

\section{Spin-0 annihilations through {$\,U$} exchanges}

The threshold behavior of the annihilation cross section
$\,\sigma_{ann}\,(\varphi\bar\varphi\! \to f\bar f)\,$
may be understood from simple arguments based on Charge Conjugation.
The initial $\,\varphi\,\bar\varphi\,$ state
has $\,C=+$ in an $S$ wave ($L=0$).
The final $\,f \bar f\,$ state then also has $\,C'=(-)^{(L'+S')} = +\ $,
\,since angular momentum conservation requires $\,J'=J=0\,$.
In the case of a {\it axial\,} coupling \,($f_A$)\, to the fermion
field $f$, the relevant terms in the Lagrangian density (\ref{effphi}),
being $\,C$-violating, cannot induce the decay
$\,\varphi\bar \varphi_{\,S\hbox{\footnotesize -wave}}\to f\bar f\,$.
For a {\it vector\,} coupling \,($f_V$),\,
the relevant terms are $\,C$-conserving,
but the $\,f\bar f\,$ final state, being vectorially produced
through the virtual production of a $U$ boson
(as if it were through a one-photon exchange),
must have $\,C'=-$
(instead of  $\,C'=+\,$ from $J$ conservation).
In both cases
there can be
no $S$-wave term in the annihilation cross section.
The dominant ($P$-wave) terms in $\,\sigma_{ann}\,v_{rel}\,$
are proportional to the square of the Dark Matter particle velocity
in the initial state,
\be
\sigma_{ann}\,v_{rel}\,(\,\varphi\bar\varphi \to f\bar f\,) \,\propto\
v_{dm}^{\,2}
\ee
at threshold\,\footnote{In other terms, the $U$ charge-density
and current of a $\,\varphi\bar\varphi\,$ pair vanish
at threshold; the annihilation amplitudes
vanish proportionally to the rest-frame momenta
$\,p_{dm}\,$ of the initial particles, or to $\,v_{dm}$,
\,as a result of the derivative nature of the $U$ coupling to scalar particles.}.

\vspace{2mm}

For a vector coupling to fermions ($f_V$),
the annihilation cross section
(which may also be obtained directly from the corresponding
production cross section in $\,e^+e^-\,$ annihilations,
cf. Sec. \ref{sec:relating}) is given by
\be
\label{sigmaannsc}
\hbox {\framebox [14cm]{\rule[-.6cm]{0cm}{1.4cm} $ \displaystyle {
\ba{c}
\displaystyle
\sigma_{ann}\,v_{rel}\,(\,\varphi\bar\varphi \to f\bar f\,)
\ \ =\ \frac{2}{3\,\pi}\ \,v_{dm}^{\ 2}\
\frac{C_U^{\,2}\,f_V^{\,2}}{(m_U^{\,2}-4\,E^2)^2}\ \
\hbox{\footnotesize$\displaystyle \sqrt{\ \hbox{\normalsize 1}-\frac{m_f^{\,2}}{E^2}}$}\ \,
{\hbox{\LARGE(}}\,E^2+\,\frac{m_f^{\,2}}{2}\,{\hbox{\LARGE)}}\ \ ,
\ea
}$}}
\ee
where
$\,\beta_f = v_f\ (/c)=\hbox{$(\ 1-m_f^{\,2}/E^2)^{1/2}$}$, and
\be
\small
\label{phasespacevector}
\frac{3}{2}\ \beta_f\ -\ \frac{1}{2}\ \beta_f^{\,3}\ \ =\ \
\hbox{\footnotesize $ \displaystyle
\sqrt{\ \hbox{\normalsize 1}-\frac{m_f^{\,2}}{E^2}}$}\ \ \,
{\hbox{\Large(}}\,E^2+\,\frac{m_f^{\,2}}{2}\,{\hbox{\Large)}}\ \ ,
\ee
is the usual kinematic factor relative to the vectorial pair production
of spin-$\frac{1}{2}$ Dirac fermions.

\vspace{2mm}

For an axial coupling ($f_A$), replacing $f_V$ by $f_A$, and (\ref{phasespacevector})
by the factor $\,\beta_f^{\,3}\,$ appropriate to the axial production
of a pair $\,f\bar f\,$ of \hbox{spin-$\frac{1}{2}$} particles, we get
\be
\ba{c}
\label{annaxial}
\displaystyle
\!\!\sigma_{ann}\,v_{rel}\,(\,\varphi\bar\varphi \to f\bar f\,)\ = \ \frac{2}{3\,\pi}\ \,v_{dm}^{\ 2}\
\frac{C_U^{\,2}\,f_A^{\,2}}{(m_U^{\,2}-4\,E^2)^2}\ E^2\
{\hbox{\LARGE(}}\,\,1-\frac{m_f^{\,2}}{E^2}\,{\hbox{\LARGE)}}^{3/2}.
\ea
\ee
If the $U$ coupling to the fermion field $f$ includes
both vector and axial contributions,
the annihilation cross section is the sum of the above
contributions (\ref{sigmaannsc}) and (\ref{annaxial}). In all cases, if behaves at threshold
like $\,v_{dm}^{\,\ 2}\,$.

\section{Spin-$\frac{1}{2}$ annihilation cross sections through {$\,U$} exchanges}

In this case also, there will be {\em no $S$-wave annihilations}, for a $U$ boson having {\em \,axial\,} couplings to Dark Matter, with {\em \,vector\,} ones to ordinary matter \cite{fay}.

\vspace{2mm}

Let us consider spin-$\frac{1}{2}\,$ Majorana fermions $\,\chi\,$, which can only have a purely
{\em axial\,} coupling to the $U$ boson. The analysis, in fact, applies as well to Dirac fermions ($\psi$), provided they are also {\em axially} coupled to the $U$ boson (i.e. excluding the presence of a vector coupling of the $U$ to the Dark Matter Dirac fermion $\,\psi$) \footnote{To
relate the annihilation cross sections of Dirac and Majorana particles
we can write the decomposition
$\,\psi=(\chi-i\chi')/\sqrt 2\ $ of
the Dirac field $\,\psi\,$,
so that $\
\bar\psi\,\gamma_\mu\gamma_5\,\psi \,=\,
\frac{1}{2}\ \bar\chi\,\gamma_\mu\gamma_5\,\chi \ +\
\frac{1}{2}\ \bar\chi'\,\gamma_\mu\gamma_5\,\chi'
$,
\,and consider an initial state
in which each of the two annihilating particles
is either  a $\,\psi\,$ or a $\,\bar\psi\,$ (or, equivalently,
either a $\,\chi\,$ or a $\,\chi'\,$).
The pair annihilation cross section
of Dirac particles ($\,\psi$, with axial coupling $\,C_U$)
is the same as for Majorana particles ($\,\chi$, with axial coupling $\,C_U/2$),
so that
$\
\label{annpsiphi2}
\sigma_{ann}\,v_{rel}\,(\,\psi\,\bar\psi\,\to\, e^+e^-\,)\ =\
\sigma_{ann}\,v_{rel}\,(\,\chi\,\chi\,\to \,e^+e^-\,)\,.
$
}.

\vspace{2mm}

The effective Lagrangian density may now be written as
\be
\label{effchi}
\hbox {\framebox [11cm]{\rule[-.5cm]{0cm}{1.2cm} $ \displaystyle {
{\cal L}\ \ =\ \ \frac{C_U}{2\ m_U^{\,2}}\ \ \
\bar\chi\,\gamma_\mu\gamma_5\,\chi\ \
(\,f_V\ \bar f\,\gamma^\mu f\ +\ f_A\ \bar f\,\gamma^\mu\gamma_5 f\,)\ \ .
}$}}
\ee
The coupling $\,f_V\,$ (i.e. in fact $\,C_Uf_V$) is now $C$-violating while $\,f_A\,$
(still generally presumed to be absent
as it would be related with an unwanted axionlike behavior of the $U$ boson, as we have discussed in subsection {\ref{subsec:limax})
would be $C$-conserving.

\vspace{2mm}

At threshold the antisymmetry of a 2-Majorana $\,\chi\,\chi$ state
imposes that the total spin be $J=S=0$,
so that the production (indifferently through vector and/or axial couplings to $f$)
of massless fermion pairs $\,f\bar f\,$
with total angular momentum $\,\lambda=\pm 1\,$ is forbidden.
But the ($S$-wave) annihilation cross section $\,\sigma_{ann}\,v_{rel}\,$
could in principle still include, at threshold (potentially dangerous) non-vanishing contributions
proportional to $\,m_f^{\,2}$ \,--\, depending on whether the $U$ coupling to $f$ is taken to be vector or axial, as we shall see.

\vspace{2mm}

The initial $\,\chi\,\chi$ state has $\,C=+\,$.
\,If it is in an $S$ wave (so that $J=S=0$), the final $\,f\bar f\,$ state must have $\,C'=(-)^{(L'+S')}=+$ \,(with $\,L'=S'$ from $J=0$).
The ($C$-violating) coupling product $\,C_U f_V\,$ cannot contribute to the $S$-wave
$\,\chi\,\chi\to f\bar f\,$ annihilation amplitude.
(Also, a $\,U\,$ with a vector coupling to $\,f\,$ can only produce
a $\,f \bar f\,$ pair with \hbox{$\,C'=-$,}
while $\,C'=+\,$ from angular momentum conservation.)
The $\,f_V\,$ contribution to the $S$-wave annihilation cross section must vanish,
so that
\be
\displaystyle
\sigma_{ann}\,v_{rel}\ (\,\chi\,\chi \to\bar f\,f\,)\ \ \propto
\hbox{\large$\displaystyle \ \ v_{dm}^{\,2}$}\ \ \
\hbox{at threshold}\ \ .
\ee
This would be different for
an {\it \,axial\,} coupling $f_A$ (corresponding to $C=+$).
The $\,f_A\,$ contribution to $S$-wave annihilation
has now no reason to vanish, as soon as
$\,m_f\neq 0\,$.
Constant terms proportional to $m_f^{\,2}\,$
(potentially undesirable for $\,m_{dm}\,\simle 100$ MeV)
\,then appear in
$\,\sigma_{ann}\,v_{rel}\,$, \,as shown later in eq.\,(\ref{sigmaannsc3}).

\vspace{2mm}

It is remarkable that the possible constraint of {\it \,vector\,} couplings of the light $U$
to the matter fermions $f$, obtained here from the requirement
of spin-$\frac{1}{2}$ annihilation cross sections strictly behaving like $\,v_{dm}^{\,2}$,
corresponds to the one following from the fact
that a light $U$ boson would have an unacceptable axionlike behavior
if it had sizeable axial couplings $f_A$ to the matter fermions $f$ \cite{pfu}, as discussed in subsection \ref{subsec:limax}.

\vspace{2mm}

As for spin-0 particles,
the annihilation cross sections of spin-$\frac{1}{2}$ Dark Matter particle pairs (\,$\chi\chi\,$ or $\,\psi\bar\psi\,$)
into $e^+e^-$ may be related to their production cross sections in $e^+e^-$ annihilations
(cf. Sec. \ref{sec:relating}).
From the relation between the pair production of (axially-coupled) \hbox{spin-$\frac{1}{2}$} particles, and spin-0 particles, in $e^+e^-$ annihilations, we find that their annihilation cross sections are the same, namely, when the electron mass is neglected,
\be
\label{annprime0}
\sigma_{ann}\,v_{rel}\,(\,\chi \,\chi \to e^+e^-)
\ \ =\ \
\frac{2}{3\,\pi}\ v_{dm}^{\,2}\
\hbox{\small $\displaystyle \frac{C_U^{\,2}\,f_{\,V}^2}{(m_U^{\,2}-4\,E^2)^2}$}
\ \ E^2\,.
\ee
Since no $\,S$-wave annihilation cross section may be induced
from a non-vanishing $m_e\,$ (in the case of $\,f_V$),
the only expected effect of a non-vanishing $\,m_e$
(or more generally of fermion masses $\,m_f$)
\,is a multiplication by the usual kinematic factor for
the vectorial production of a $\,f\bar f\,$ pair,
$\,\frac{3}{2}\ \beta_f\,-\,\frac{1}{2}\ \beta_f^{\,3}\,$:
\be
\label{sigmaann12}
\hbox {\framebox [14cm]{\rule[-.5cm]{0cm}{1.4cm} $ \displaystyle {
\sigma_{ann}\,v_{rel}\,(\,\chi \,\chi \to f \bar f)
\ \ =\ \
\displaystyle
\frac{2}{3\,\pi}\ \,v_{dm}^{\ 2}\
\frac{C_U^{\,2}\,f_V^{\,2}}{(m_U^{\,2}-4\,E^2)^2}\ \
\hbox{\small$\displaystyle \sqrt{\ 1-\frac{m_f^{\,2}}{E^2}}$}\ \,
{\hbox{\Large(}}\,E^2+\,\frac{m_f^{\,2}}{2}\,{\hbox{\Large)}}\ \ ,
}$}}
\ee
exactly as in (\ref{sigmaannsc}) for the pair annihilation of spin-0 particles.
The same formula also applies for the annihilation cross section of Dirac particles,
$\,\sigma_{ann}\,v_{rel}\,(\,\psi \,\bar\psi \to f \bar f)\,$,
\,as already mentioned.

\vspace{2mm}
Remarkably enough, this cross section (\ref{sigmaann12})
for the annihilation of spin-$\frac{1}{2}$
particles (axially coupled to the $U$ boson, this one vectorially coupled to the matter fermion $f$)
is identical to (\ref{sigmaannsc})
for the annihilation of spin-0 candidates\,!
We get in both cases the same $\,v_{dm}^{\,2}\,$ suppression factor,
as desirable to avoid an excessive production of gamma rays
from residual light Dark Matter annihilations.
The (collisional and free-streaming) damping effects  \cite{class,bf,damp} associated
with such particles are also, in both cases, sufficiently small.

\vspace{2mm}

For comparison, the spin-$\frac{1}{2}$ annihilation
cross section in the case of an axial matter fermion coupling $\,f_A\,$ (still with an axial coupling of the Dark Matter fermion $\chi$ or $\psi$ to the $U$, and assuming for simplicity the $U$ heavy compared to $\,m_{dm}\,$ and $m_f$) is:
\be
\label{sigmaannsc3}
\hbox {\framebox [15.8cm]{\rule[-.55cm]{0cm}{1.4cm} $ \displaystyle {
\sigma_{ann}\,v_{rel}\,(\,\chi \,\chi \to f \bar f)\ =\ \hbox{$\small\displaystyle \frac{1}{2\,\pi}\ \,
\frac{C_U^{\,2}\,f_{A}^2 }{(m_U^{\,2}\,-\,4\,E^2)^2}$}\ \
\hbox{$\sqrt{\,1-\frac{m_f^2}{E^2}}$}\ \ \normalsize
\left[\ \,\frac{4}{3}\ \,(E^2-m_f^2)\ v_{dm}^{\,2}  \,+\,
\hbox{\large $ \frac{m_{dm}^{\,2}}{E^2}  $}
\ m_f^2\ \right]\,.
}$}}
\ee

\vspace{2mm}
\noindent
It coincides with (\ref{sigmaann12})
(replacing $\,f_A\,$ by $\,f_V$) in the limit of vanishing $\,m_f$.
But the overall $\,v_{dm}^{\,2}\,$ suppression factor
no longer subsists for $m_f \neq 0$,
for which one recovers a non-vanishing $S$-wave term
in the annihilation cross section, proportional to $\,m_f^{\,2}$.

\vspace{2mm}
Such a behaviour, however, could still be tolerated, from the point of view of the annihilation cross section, if $m_{dm}$ is sufficiently large compared to $m_e$, so that the ``unwanted''
term not proportional to $\,v_{dm}^{\,\ 2}\,$,
\,damped by a factor $\,\propto\,m_e^{\,2}/m_{dm}^{\ 2}$,
\,be sufficiently small. This provides a way to ease out the
requirement of a cross section $ \,\sigma_{ann}v_{rel}\,$ proportional to $\,v_{dm}^{\ 2}$. As an example for
$\,m_{dm}\simge 10$ MeV, the coefficient of the constant term in (\ref{sigmaannsc3}),
although non-vanishing, is smaller than the coefficient of the $\,v_{dm}^{\ 2}$ term
by a factor of about $\,3\,m_e^{\,2}/(64\,m_{dm}^{\,2}) \simle 1.2\ 10^{-4}$.
The constant term in the annihilation cross section $ \,< \sigma_{ann}v_{rel}>\,$
\,-- the one potentially dangerous when considering residual annihilations
of relic low-velocity Dark Matter particles --\,
then represents less than $10^{-3}$ times the $ \,< \sigma_{ann}v_{rel}>\,$ at freeze out
(corresponding to $v_{dm}\simeq \,.4\,c$),
determined by the relic abundance estimate of Sec. \ref{sec:relic}. Remembering also that
a certain fraction of Dark Matter annihilations, possibly around 60\% or even more depending on $U$ couplings, could be into unobserved $\,\nu\bar\nu$ pairs rather than $e^+e^-$, this may well be sufficient to keep the
gamma ray production from residual Dark Matter annihilations at a sufficiently small level, also in this case of an axial-axial interaction of a \,($\simge 10$ MeV)\, Dark Matter particle with
ordinary matter.

\section{Relating production and annihilation cross sections}
\label{sec:relating}

The production (in $\,e^+e^-$ scatterings) and annihilation cross sections
of Dark Matter particles may be related using  $CPT$ (or simply $T$) invariance. One gets for example, in the case of spin-0 particles:
\be
\label{prodann}
\sigma_{prod}\,v_{e}\,(e^+e^-\!\to\varphi\,\bar\varphi)\,/v_{dm}
\ \equiv\
\frac{1}{4}\
\sigma_{ann}\,v_{dm}\,(\varphi\,\bar\varphi \to e^+e^-)\,/v_e\,,
\ee
which may also be used to derive the annihilation cross section
from the production one.
Indeed, from the electromagnetic pair production cross section
of charged \hbox{spin-0} particles in $\,e^+e^-\,$ annihilations
(neglecting $\,m_e$),
$
\,\sigma_{prod}^{\ (\gamma)}
\,=\,\frac{e^4}{48\ \pi\ s}\ \beta_{dm}^{\,3}\, $,
\,we get
\,the production cross section, through $U$ exchanges,
of neutral spin-0 Dark Matter particles,
\be
\label{sigmaprod}
\sigma_{prod}\,(e^+e^-\!\to\varphi\,\bar\varphi)\ =\
\frac{1}{12\,\pi}\ \frac{C_U^{\,2}\,f_{\,V}^2}{(4\,E^2-m_U^{\,2})^2}\
 \ E^{2}\
\beta_{dm}^{\,3}\ \,.
\ee
Multiplying it by $\,v_{rel}\simeq 2$,
\,by the spin factor \,4\, and the velocity ratio
$\,(\beta_e\simeq 1)/(\beta_{dm}=v_{dm}),\,$
we get the corresponding annihilation cross section,
\be
\label{ann}
\sigma_{ann}\,v_{rel}\, (\,\varphi\,\bar\varphi\to\,e^+e^-\,)\ =\
\frac{2}{3\,\pi}\ v_{dm}^{\,2}\
\frac{C_U^{\,2}\,f_{\,V}^2}{(m_U^{\,2}-4\,E^2)^2}\ \ E^2\ ,
\ee
which, once the kinematic factor
$\,\frac{3}{2}\,\beta_f-\frac{1}{2}\,\beta_f^{\,3}$
\,is reintroduced, coincides with (\ref{sigmaannsc}).
(Replacing $\,f_V$ by $\,f_A\,$, and reintroducing the
kinematic factor $\,\beta_f^{\,3}$, \,we recover expression (\ref{annaxial})
of the annihilation cross section through an axial coupling to $f$.)

\vspace{2.5mm}
The $\,v_{dm}^{\,2}\,$ suppression factor in the annihilation cross section
$\,\sigma_{ann}\,v_{rel}\,$ of spin-0 particles $\,\varphi\,$
appears simply as {\it a reflection by $\,CPT\,$ of the
well-known $\,\beta^3\,$ factor\,}
for their pair production in $\,e^+e^-$ annihilations
(with, in both cases, a $P$ wave for the $\,\varphi\,\bar\varphi\,$ state).
This also applies to spin-$\frac{1}{2}\,$ particles, whether Majorana or Dirac,
axially coupled to the $U$ boson.
Their pair production through an axial coupling to the $U$ also involves
a $\,\beta_{dm}^{\,3}\,$ factor,
which reflects (by $CPT$) in a $\,v_{dm}^{\,2}\,$ suppression factor
for the annihilation cross section at threshold.

\vspace{2mm}

Because the {\it \,production\,} cross sections
of spin-0 and \hbox{spin-$\frac{1}{2}$} particles in $\,e^+e^-\,$ annihilations
are given by similar formulas
the corresponding {\it \,annihilation\,} cross sections
into $\,f\bar f\,$ pairs are given, also, by similar formulas.
For spin-$\frac{1}{2}\,$ Dirac particles, eq.\,(\ref{prodann}) is replaced by
\be
\label{prodann7}
\sigma_{prod}\,v_{e}\,(e^+e^-\!\to\psi\,\bar\psi)\,/v_{dm}
\ \equiv\
\sigma_{ann}\,v_{dm}\,(\psi\,\bar\psi \to e^+e^-)\,/v_e\ \ ,
\ee
with, for an axially coupled  $U$ boson (with axial coupling $C_U$) and neglecting $m_e$,
\be
\sigma_{prod}\,(e^+e^-\!\to\varphi\,\bar\varphi)\ =\ \frac{1}{4}\
\sigma_{prod}\,(e^+e^-\!\to\psi\,\bar\psi)\ \ ,
\ee
both cross sections being proportional to $\,\beta_{dm}^{\,3}\,$,
\,so that
\be
\label{annpsiphi}
\sigma_{ann}\,(\,\psi\,\bar\psi\,\to\,e^+e^-\,)\ \ =\ \
\sigma_{ann}\,(\,\varphi\,\bar\varphi\,\to\,e^+e^-\,)\ \ ,
\ee
as given by (\ref{ann}), the same formula applying to a Majorana fermion $\chi$ (with axial coupling $\frac{C_U}{2}$). This leads to eqs.\,(\ref{sigmaannsc}), (\ref{annaxial})
and (\ref{sigmaann12}), when the appropriate phase space factors are reestablished.

\section{Constraints on the $U$ couplings}
\label{sec:value}

To get an idea of the size of the couplings
required to get appropriate values of the annihilation cross sections at freeze-out
($\approx$ \,8\,--\,10 picobarns for a complex spin-0 particle,
or \,4\,--\,5\, for a Majorana one,
\,cf.~\,Sec. \ref{sec:relic}), we can write
the above expressions (\ref{sigmaannsc},\ref{annaxial},\ref{sigmaann12}) as
\vspace{1mm}
\be
\label{sigmanum}
\hbox {\framebox [11cm]{\rule[-.7cm]{0cm}{1.6cm} $ \displaystyle {
\sigma_{ann} v_{rel}\ \simeq\
\hbox{$\displaystyle
\frac{v_{dm}^{\,2}}{.16}
\
\left(\frac{C_U\,f_{V,A}}{10^{-6}}\right)^2
\
\left(\ \frac{m_{dm}\times 3.6\ \rm{MeV}}{m_U^{\,2}-4\,m_{dm}^{\,2}}\ \right)^2
$}
\,\hbox{pb}\,,
}$}}
\ee

\noindent
to be multiplied by the kinematic
factor relative to the vectorial ($\,\frac{3}{2}\,\beta_f-\frac{1}{2}\,\beta_f^{\,3}$)
\,or axial ($\beta_f^{\,3}$) \,production of spin-$\frac{1}{2}$ $f\bar f$ pairs.
(Note that only couplings to electrons and neutrinos play an essential r\^ole in
Light Dark Matter annihilations into $e^+e^-$ or $\,\nu\bar \nu$.)
\,As discussed in subsection \ref{subsec:limax},
we consider preferentially vectorial couplings of the $\,U$ to ordinary matter ($f_{V}$),
with values much smaller than the electric charge \hbox{($e\simeq .3$)}
\,by several orders of magnitude.
The resulting $\,U$ boson effects on ordinary particle physics processes,
charged lepton $g-2$, \,etc., then appear sufficiently small \cite{bf,fay}.

\vspace{2mm}

In particular, for a vectorially coupled $\,U$ somewhat heavier than the electron
but lighter than the muon, the comparison between the additional $U$ contributions
to the $\mu$ and $e$  anomalies and the possible difference between the experimental and Standard Model values
indicates that~\cite{bf,g-2}
\be
\label{amu}
\ba{ccccc}
\delta a_\mu &\simeq &
\displaystyle
\frac{f_{V\mu}^{\,2}}{8\,\pi^2} &\simeq& (\,2\pm2\,)\ 10^{-9}
\ \ ,
\vspace{2mm}\\
\delta a_e &\simeq & \displaystyle
\frac{f_{Ve}^{\,2}}{12\,\pi^2}\ \frac{m_e^{\,2}}{m_U^{\,2}}
&\simeq& \,(\,4\pm 3\,)\ 10^{-11}\ \ ,
\ea
\ee
so that
\be
\label{amu2}
f_{V\mu}\simle \,6\ 10^{-4}\ \ ,\ \ \ \ f_{Ve}\simle \,2\ 10^{-4}\ m_U \hbox{(MeV)}\ \ .
\ee

\vspace{1mm}
\noindent
One should also have, for a $U$ mass larger than a few MeV's,
\be
|f_{V\nu}\,f_{Ve}|\ \simle\ G_F\ m_U^{\,2}\ \simeq\ 10^{-11}\ (m_U(\hbox{MeV}))^2\ \
\ee
so that $U$ exchanges do not modify excessively $\,\nu\,-\,e\,$ low-energy elastic scattering
cross sections~\cite{nu}.
This requires that the $U$ couplings to neutrinos, at least, be
sufficiently small. Conversely, the $U$ coupling $C_U$ of Dark Matter should not be
too small, so as to get, from eq.\,(\ref{sigmanum}),
large enough values of the annihilation cross section
$\,<\sigma_{ann} v_{rel}>\,$.

\vspace{2mm}

If, in addition, all $\,f_V$'s turn out to be of the same order,
they should then be smaller than about $\,3\ 10^{-6}\, m_U$(MeV), or in other terms, approximately,
\be
\frac{f_V^{\,2}}{m_U^{\,2}}\ <\ G_F\ \
\ee
(to be contrasted with eqs.~(\ref{limabc},\ref{limaq},\ref{limas}) in subsection \ref{subsec:limax},
imposing a stricter bound for an axial coupling to quarks).
Expression (\ref{sigmanum}) of
$\,<\!\sigma_{ann}v_{rel}\!>\,$  then implies
that the $U$ coupling to Dark Matter $C_U$ should be {\it \,larger\,}
 than about $\,.2\ (m_U^{\,2}-4\,m_{dm}^{\,2})/m_U\,m_{dm}$ (for $\,<\!\sigma_{ann}v_{rel}\!>\,$
to be larger than 4 or 5 pb),
in which case  {\it \,the self-interactions of Dark Matter would be quite significant},
\,much stronger than ordinary weak interactions, by several orders of magnitude
(depending also on the energy considered)!

\vspace{2mm}

When the $U$ and $\,\varphi\,$ particles are light (and neglecting $m_e$),
eq. (\ref{sigmaprod}) may be written as
\be
\hbox {\framebox [15cm]{\rule[-.6cm]{0cm}{1.4cm} $ \displaystyle {
\sigma_{prod}\,(\,e^+e^-\to\varphi\,\bar\varphi\,) \ \simeq\
\hbox{\small $\displaystyle
\frac{C_U^{\,2}\,f_{\,V}^2}{48\,\pi\ s}$}\
\simeq \ \displaystyle{ C_U^{\,2}\,f_{V}^{\,2}\ \
\frac{2.58\ \mu\rm{b}}{s\,(\hbox{\footnotesize GeV}^2)}
}
\ \simeq \ \left(\frac{C_U\,f_V}{10^{-6}}\right)^2\ \
\frac{2.6\ 10^{-42}\ \hbox{cm}^2}
{\left(\sqrt s\,(\hbox{\footnotesize GeV})\right)^2}\ ,
}$}}
\ee
which may also be applied in the case of an axial coupling by changing $\,f_V$ into $\,f_A$.
This should still be multiplied by 4, or 2, for the production
of a spin-$\frac{1}{2}$ $\,\psi\bar\psi\,$ or $\,\chi\chi\,$ pair, respectively.
It is easy to get from there the corresponding cross sections for
$\,e^+e^-\to \gamma \ +$ a pair of unobserved Dark Matter particles, and to compare them (as in \,\cite{sigmaphot}) to the cross sections for $\,e^+e^-\to\,\gamma\ \nu\,\bar\nu\,$ or
$\,e^+e^-\to\,\gamma\ \tilde \gamma\,\tilde\gamma\,$.
For the relevant values
of $\,C_U f_V$ \,considered,
these production cross sections get very small at high energies, much below
neutrino production cross sections,
so that the direct production of such Dark Matter particles (and $U$ bosons as well)
is in general not expected to lead to easily observable signals in
$e^+e^-$ annihilations, although this would deserve further studies \cite{bf}.
Note that, depending on the relative values of $\,C_U$ and $f_{Ve}$
(and to a lesser extent $f_{V\nu}$),
especially if we take $C_U$ relatively large compared to $f_V$, as indicated earlier,
the $U$ boson may in general be expected to have preferred invisible decay modes into Dark Matter particles,
dominating over visible decays $\,U\,\to\,e^+e^-$.


\section{Final remarks}

Altogether
spin-$\frac{1}{2}$ Dark Matter particles {\it axially coupled} to the $U$ boson have
the required characteristics for
Light Dark Matter (LDM) particles annihilating into $\,e^+e^-$,
\,as well as spin-0 particles.
In both cases, $U$-induced Dark-Matter/electron interactions
should be significantly stronger than ordinary weak interactions
at low energy (but weaker at high energies),
which requires the $U$ to be more strongly coupled to Dark Matter
than to ordinary matter
\,-- also resulting in significant $\,U$-induced Dark Matter self-interactions.
Finally, light spin-$\frac{1}{2}$ Dark Matter particles
appear more attractive than \hbox{spin-0} ones,
as the smallness of their mass is easier to understand, and
provide valuable alternative scenarios to be discussed and confronted with the standard ones.

\end{document}